# Technical Report

*Building Brain Invaders: EEG data of an experimental validation*

~

GFP. Van Veen, A. Barachant, A. Andreev, G. Cattan, P. Rodrigues, M. Congedo

GIPSA-lab, CNRS, University Grenoble-Alpes, Grenoble INP.
Address : GIPSA-lab, 11 rue des Mathématiques, Grenoble Campus BP46, F-38402, France

**Abstract** - We describe the experimental procedures for a dataset that we have made publicly available at https://doi.org/10.5281/zenodo.2649006 in *mat* and *csv* formats. This dataset contains electroencephalographic (EEG) recordings of 25 subjects testing the *Brain Invaders* (1), a visual P300 Brain-Computer Interface inspired by the famous vintage video game *Space Invaders* (Taito, Tokyo, Japan). The visual P300 is an event-related potential elicited by a visual stimulation, peaking 240-600 ms after stimulus onset. EEG data were recorded by 16 electrodes in an experiment that took place in the GIPSA-lab, Grenoble, France, in 2012 (2,3). Python code for manipulating the data is available at https://github.com/plcrodrigues/py.BI.EEG.2012-GIPSA. The ID of this dataset is *BI.EEG.2012-GIPSA*.

**Résumé** - Dans ce document, nous décrivons une expérimentation dont les données ont été publiées à https://doi.org/10.5281/zenodo.2649006 aux formats *mat* et *csv*. Ce jeu de donnée contient les enregistrements électroencéphalographiques (EEG) de 25 sujets testant *Brain Invaders* (1), une interface cerveau-ordinateur de type 'P300 visuel' inspirée du fameux jeu vintage *Space Invaders* (Taito, Tokyo, Japan). Le P300 visuel est une perturbation du signal EEG apparaissant 240-600 ms après le début d'une stimulation visuelle. L'EEG de chaque sujet a été enregistré grâce à 16 électrodes réparties sur la surface du scalp. L'expérience a été menée au GIPSA-lab (Université de Grenoble-Alpes, CNRS, Grenoble-INP) en 2012 (2,3). Nous fournissons également une implémentation python pour manipuler les données à https://github.com/plcrodrigues/py.BI.EEG.2012-GIPSA. L'identifiant de cette base de données est *BI.EEG.2012-GIPSA*.

**Introduction**

The visual P300 is an event-related potential (ERP) elicited by a visual stimulation, peaking 240-600 ms after stimulus onset. The experiment was designed to validate the game design of *Brain Invaders* (1), a visual P300 Brain-Computer Interface inspired by the famous vintage video game *Space Invaders* (Taito, Tokyo, Japan). The Brain Invaders is based on a P300-based brain-computer interface (BCI) working on a PC. In (1,2) these data were classified using the xDAWN spatial filter (4). This experiment features a training-test mode of operation and both a longitudinal and transversal design. An example of applications of this dataset can be seen in https://github.com/plcrodrigues/py.BI.EEG.2013-GIPSA. Other datasets of Brain Invaders experiment are presented in (5,6). For example, reference (5) features both a training-test (classical) mode of operation and a calibration-less mode of operation, as described in (7–9). This is the first experiment ever carried out using the Brain Invaders. The complete list of experiments is available at https://sites.google.com/site/marcocongedo/science/eeg-data.

**Participants**

26 subjects participated in the experiment (7 females), with mean (sd) age 24.4 (2.76). The youngest subject was 21 and the oldest 31. One subject was excluded from the study due to material issues during the experiment. Half of them played games occasionally, that is, around 4.5 hours a week. All subjects were volunteers recruited by means of flyers and of the mailing list of the University of Grenoble-Alpes. All participants provided written informed consent confirming the notification of the experimental process, the data management procedures and the right to withdraw from the experiment at any moment.

**Material**

EEG signals were acquired by means of the NeXus-32 (MindMedia, Herten, Germany), a research-grade amplifier and EEG headset. The cap was equipped with 16 Silver/Silver Chloride wet electrodes, placed according to the 10-20 international system (**Figure 1**). The locations of the electrodes were F7, F3, F4, F8, T7, C3, CZ, C4, T8, P7, P3, PZ, P4, P8, O1 and O2. The ground was placed at the FZ scalp location. The NeXus-32 machine is produced by Twente Medical Systems International B.V. (TMSi, Enschede, The Netherlands). It does not use an electrode as reference, rather, a, hardware common average reference is used. The amplifier was linked by USB connection to the PC where the data were acquired by means of the software OpenVibe (10,11). Data were acquired at a sampling frequency of 128 samples per second. For ensuing analysis, the application tagged the EEG using *software tagging*. The tags were sent by the application to the OpenVibe plateform thanks to the Boost inter-process messaging (12). Note that the tagging process introduces a jitter and a latency which artificially modify the ERPs onset. These belong to the hardware and software components of the experiment. In particular, a disadvantage of software tagging is a strong drift over time, resulting in higher jitter (2,13). As a consequence, it is only possible to compare the ERP acquired within the same experimental conditions when the latency is not corrected (14).

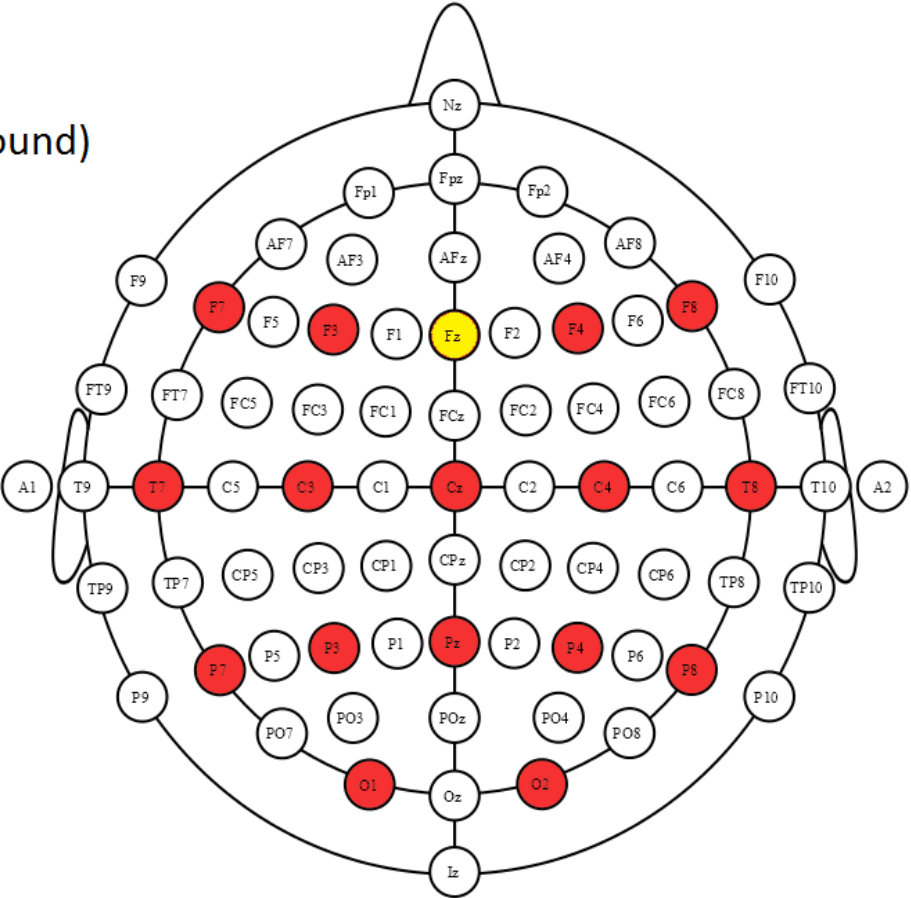

**Figure 1.** In red, the 16 electrodes placed according to the 10-20 international system. We used FZ (in yellow) as ground.

**Procedures**

The interface of Brain Invaders is composed of 36 aliens (symbols) that flashes on the computer screen in 12 groups of six aliens. In the Brain Invaders P300 paradigm, a *repetition* is composed of 12 flashes (*i.e.*, one for each group), of which two include the Target symbol (*Target* flashes) and 10 do not (*non-Target* flash) - **Figure 2**. At the end of each repetition each alien has flashed exactly twice. The number of Target and non-Target flashed are therefore in a ratio one-to-five. The total number of Target and non-Target flashes is variable across sessions since the number of repetitions needed to destroy the Target in the Brain Invaders BCI video game depends on the user's performance (7,8). Please see (1,13) for a full description of this paradigm. In any case, since the classes are unbalanced, an appropriate score must be used for quantifying the performance of classification methods, such as the Balanced Accuracy (BA):

$$BA = \frac{1}{2}(\frac{A}{A+B} + \frac{C}{C+D}),$$

where A and B (resp. C and D) stands for the number of correctly and non-correctly classified flashes of non-Target (resp. Target) group.

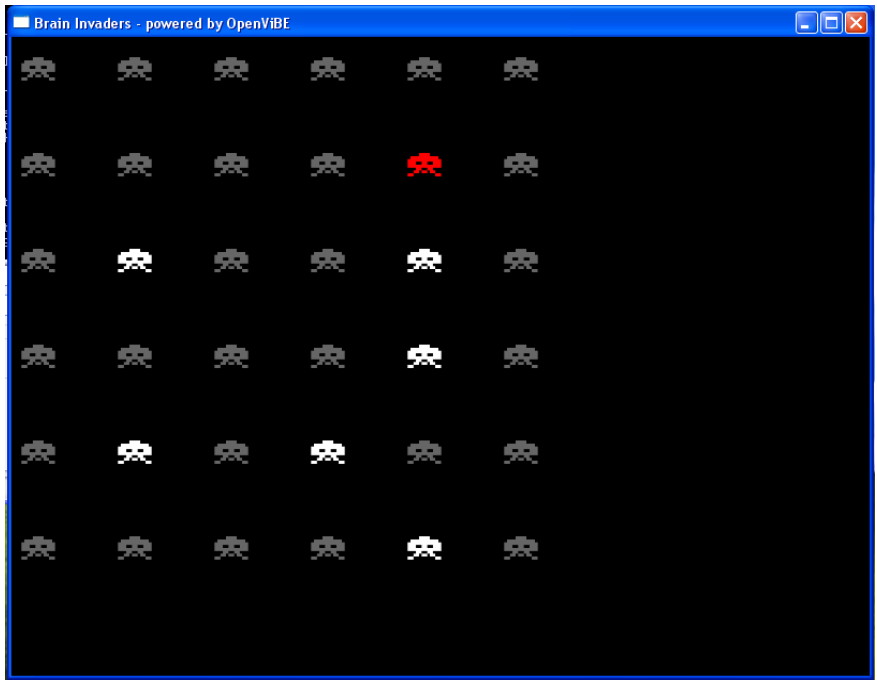

**Figure 2.** Interface of Brain Invaders at the moment where a group of six non-Target symbols flash (in white). The red symbol is the Target. The non-Targets that are not flashing are in grey.

For all subjects, the experiment took place in a small room of four-m² surface, containing the PC screen and all the required hardware materials for acquiring the EEG data. The subject was sitting at a distance of 75 to 115 cm from the screen. The EEG headset was placed on all subjects, and the integrity of the whole recording pipeline was checked by performing preliminary tests. The experimenter controlled the session from an adjacent room equipped with a one-way glass window.

Before experiment onset, the subjects were instructed to limit eye blinks, head movements and face muscular contractions, which disrupt the EEG signal. In order to help the subjects maintaining concentration, the subjects were asked to silently count the number of Target flashes.

Each subject participated in a *Training* and *Online* session. In the training session the Target alien was chosen randomly at each repetition through the use of a predefined randomised list. The task of the user was to focus on the Target alien. Once height repetitions of flashes were completed, a new Target was selected until the Target list was completed. There were eight Targets in this list, resulting in 128 Target trials (8 Targets x 8 Repetitions x 2 Flashes) and 640 non-Target flashes. Note that this number may slightly vary as it happened that the recording accidentally stopped before the end of a session. After this was done, the BCI was calibrated.

The player's task was the same on the Online and Training session, although the visual interface was different. In the online session, distractors were introduced in the game. The Online session consisted of three levels. In the first level a group of six by six aliens moved from left to right across the screen. The Target alien was placed roughly in the middle of the screen. Besides the Target there were only regular typed aliens. Level 1 can be seen in **Figure 3a**. Level 2 was similar to the first level, where again a group of six by six aliens moved from left to right across the screen. However, the Target was placed near the bottom left corner and had four distractors, two of each type, placed next to it diagonally. Level 2 can be seen in **Figure 3b**. The third and final level was more complicated than the first two. The Target alien was placed in the centre, next to three additional red typed distractors. The Target and distractors moved in a circle around the centre of the screen. Both on the left and on the right of the Target alien a group of blue distractors was placed. They moved around an elliptical trajectory. The remainder of the grid was filled in with regularly typed aliens. Level 3 can be seen in **Figure 3c**.

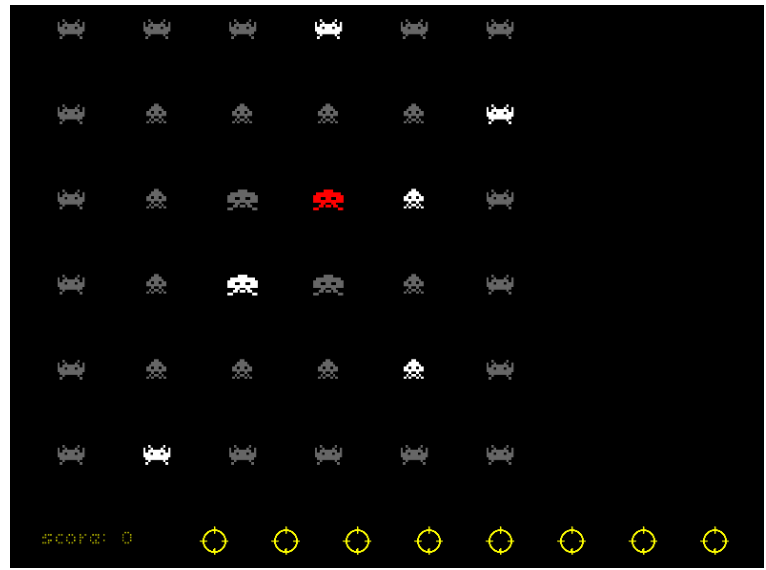

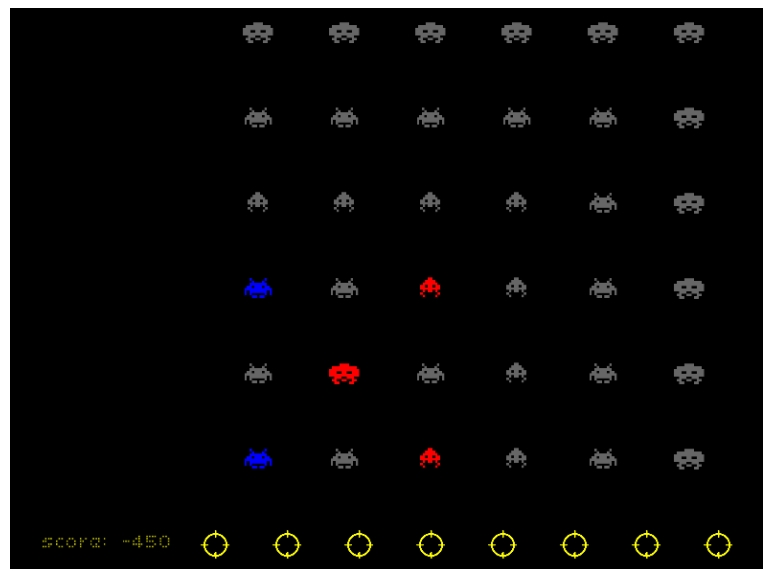

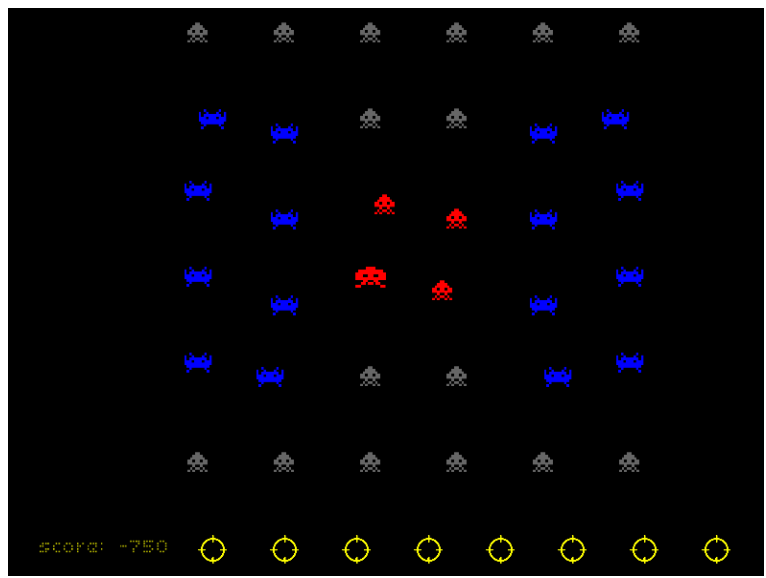

**Figure 3.** Screenshots of the level 1 (a), 2 (b) and 3 (c).

The player played all levels for a minimum of three and a half minutes. To be more precise, rather than loading the next level when the Target had been hit, the level was repeated until the minimum time had passed; after three-and-an-half-minutes had elapsed, the level would no longer be reload once the Target was hit. We choose this timeframe because it results in a minimum of four rounds in the worst-case scenario. In order to counter-balance learning and difficulty effects, half of the players played the levels in reverse order (so they played the level estimated most difficult first and the one deemed easier at the end of the experiment, instead of vice versa).

**Organisation of the Dataset**

For each subject we provide two *mat* (MathWorks, Natick, US) and *csv* files containing the complete recording of the Training and Online sessions. Each file is a 2D matrix where the rows contain the observations at each time sample. Columns 2 to 18 contain the recordings on each of the 17 EEG channels (16 electrodes plus a ground) which order is reported in **Figure 1**. The first column of the matrix represents the timestamp of each observation. The rows of column 19 (Flash) are filled with zeros, except at the timestamp corresponding to a stimulation onset. The rows of column 20 (Target) are filled with zeros, except at the timestamp corresponding to the flash onset of a Target symbol.

We supply an online and open-source example working with Python (15) and using the analysis framework MNE (16,17) and MOABB (18,19), a comprehensive benchmark framework for testing popular BCI classification algorithms. This example shows how to download the data and classify 1s non-Target and Target epochs of signals. This database has been used in the development of *Brain Invaders 2* (20), a complete state-of-the-art P300-based BCI chain for single and multi-users.